\documentclass[twocolumn,showpacs,preprintnumbers,amsmath,amssymb,floatfix]{revtex4}

\usepackage[dvips]{graphicx}% Include figure files
\usepackage{dcolumn}% Align table columns on decimal point
\usepackage{bm}% bold math

\def\beq{\begin{equation}}
\def\eeq{\end{equation}}

\begin{document}

\preprint{}

\title{Topological constraints on magnetostatic traps}

\author{R. Gerritsma} \email{gerritsm@science.uva.nl} \author{R.~J.~C. Spreeuw}
\affiliation{Van der Waals-Zeeman Institute, University of Amsterdam,\\
Valckenierstraat 65, 1018 XE Amsterdam, The Netherlands}
\homepage{http://www.science.uva.nl/research/aplp/}

\date{\today}% It is always \today, today,

\begin{abstract} 
We theoretically investigate properties of magnetostatic traps for cold atoms 
that are subject to externally applied uniform fields. We show that Ioffe Pritchard traps and other 
stationary points of $B$ are confined to a two-dimensional curved surface, or manifold $\cal{M}$, defined by 
$\det(\partial B_i/\partial x_j)=0$. 
We describe how stationary points can be moved over the manifold by applying external uniform fields. 
The manifold also plays an important role in the behavior of points of zero field. 
Field zeroes occur in two distinct types, in separate regions of space divided by the manifold. Pairs of 
zeroes of opposite type can be created or annihilated on the manifold. 
Finally, we give examples of the manifold for cases of practical interest.
\end{abstract}


%\keywords{}

\maketitle

\section{Introduction}

Magnetic trapping of neutral particles has first been achieved for cold neutrons \cite{KugPauTri78} and has since become a widely used tool in cold-atom physics \cite{MigProMet85}. 
More recently, the flexibility to design complicated magnetic trapping potentials 
has been boosted tremendously by the development of atom chips \cite{FolKruHen02,Rei02,DekLeeLor00}. The field sources are defined by microfabrication on a planar substrate, taking the form of either current-carrying wires or patterns in a permanent magnetic film \cite{SinCurHin05,HalWhiSid06,BarGerSpr05}. 

Magnetostatic traps are defined by magnetic field minima. 
In this paper we investigate the occurrence of field minima from a general perspective  \cite{Dav02}. 
We introduce a novel conceptual tool, a curved surface, or manifold $\cal{M}$, to which all stationary points of
$B$ (non-zero minima and saddle points) are confined. 
We derive expressions for the movement of stationary points over this manifold, in response to a change of
an external uniform control field. 
We also show that the same manifold plays an important role in the creation and merging of field zeroes.

The typical application that we have in mind is a situation where a magnetic field configuration
is fixed by e.g. permanent magnets and control of the field is limited to the application of
uniform external fields. This situation occurs for instance in atom chip experiments \cite{SinCurHin05,HalWhiSid06,BarGerSpr05}, where
field gradients can become very large. Control of the movement of Ioffe-Pritchard (IP)
traps \cite{GotIofTel62,pri83} is of importance in loading procedures and in experiments that require dynamical
splitting or movement of atomic clouds. During loading for instance it is important
to avoid regions of zero field, since this will lead to losses due
to Majorana spin flips to untrapped states. It is also important to avoid unwanted
splitting of the trap during the transport and compression of the cloud to the final
trap. Furthermore, quantum information processing applications on an atom chip may
require the movement of qubits to regions where they can be `read out' or manipulated. 
In this case it is also of importance to keep track of the individual phase evolutions
of the atoms, i.e. to control the trapping parameters during transport.

Although we investigate here magnetic traps, most of our conclusions also
apply to traps based on electrostatic fields insofar as they rely on the field
being rotation and divergence free. Electrostatic traps can be used to trap molecules
with an electric dipole moment \cite{BetBerMei00, CroBetMei01, RieJunRem05}.

This paper is structured as follows. After introducing our notation in Sec.~\ref{sec:Notation},
in Sec.~\ref{sec:StationaryPoints} we derive the expression for the manifold to which stationary points must be 
confined. We also show how to create a IP trap in a given point on this manifold. In Sec.~\ref{sec:MovingFieldMinima} 
we derive an expression for the movement of stationary points along the manifold, under the influence
of an external uniform field. In Sec.~\ref{sec:TheManifoldAndFieldZeroes} we investigate the relationship between the manifold and points of zero field.
We show how field zeroes can be moved and how pairs of zeroes can be created and annihilated on the manifold.
Finally, in Sec.~\ref{sec:practical} we investigate the shape of the manifold for some cases of experimental interest.

\section{Notation}
\label{sec:Notation}

Magnetic traps are usually operated in the regime where moving particles experience a magnetic field that varies slowly compared to the Larmor spin precession frequency. The spin component parallel to the field is then conserved due to adiabatic following of the local direction of the magnetic field. The effective potential is proportional to the modulus of the field: 
$B(\mathbf{r})=|\mathbf{B}(\mathbf{r})|$.
We are interested in stationary points and trapping
frequencies in this potential. For this purpose it is equivalent to use
$B^2(\mathbf{r})$, since a minimum or saddle point of $B$ is also a
minimum or saddle point of $B^2$. We shall use
$B^2(\mathbf{r})$ for convenience and define 
\beq
U(\mathbf{r}) = B^2(\mathbf{r}).
\eeq
Throughout this paper we adopt the convention that
summation over repeated indices is implied, e.g. $B_i B_i\equiv
B_x^2+B_y^2+B_z^2=B^2$. Points where $U$ is stationary are defined by
$\partial_i U\equiv\partial U/\partial x_i=0$ for all $i=1,2,3$. In
order to decide whether a stationary point is a local minimum or a
saddle point, we will need also the second derivatives. Therefore, let
us expand $\mathbf{B}(\mathbf{r})$ to second order in the relative coordinates
$x_i$ around some point of interest, 
\beq
B_i=u_i+g_{ij}x_j+\frac{1}{2}c_{ijk}x_j x_k+ \mbox{h.o.t.} \label{Bexpand}
\eeq
where h.o.t. denotes higher order terms, $g_{ij}=\partial_j B_i$ is a
tensor describing the gradient of the vector field $\mathbf{B}(\mathbf{r})$ and
$c_{ijk}=c_{ikj}=\partial_j\partial_k B_i$ is a curvature tensor. 

Using Maxwell's equations for stationary fields in vacuum we can
impose some restrictions on the tensor components $g_{ij}$ and
$c_{ijk}$. 
From the conditions $\mbox{div}\,\mathbf{B}=0$ and
$\mbox{curl}\,\mathbf{B}=0$ for stationary fields in empty space we see
that the gradient tensor must be both traceless and symmetric, 
\begin{eqnarray}
g_{ii} & = & 0 \\
g_{ij} & = & g_{ji}
\end{eqnarray}
This leaves five independent parameters for $g_{ij}$, which can be
interpreted as follows. In the coordinate frame where $g_{ij}$ is
diagonal, two independent gradients can be chosen. Three angles are
needed to specify the orientation of the coordinate frame. 

Similarly, the curvature tensor $c_{ijk}$ must be fully
symmetric under permutation of indices and all its partial traces must vanish:
{\setlength\arraycolsep{2pt}
\begin{eqnarray}
c_{iij} & = & c_{iji} = c_{jii} = 0 \label{zeropartialtrace} \\ 
c_{ijk} & = & c_{ikj} = c_{kji} 
\end{eqnarray}}
This leaves seven independent parameters for $c_{ijk}$. 

Throughout the paper we adopt the convention that eigenvalues of
a tensor are written in capital letters and eigenvectors are written
in capital bold face letters. Thus $(\mathbf{G_1},\mathbf{G_2},\mathbf{G_3})$
are the eigenvectors of $g_{ij}$ and $(G_1,G_2,G_3)$ are the corresponding
eigenvalues.

\section{Stationary points}
\label{sec:StationaryPoints}

\subsection{The manifold $\mbox{det}(g_{ij})=0$}

In order to find the stationary points of $U=B_i B_i$, we substitute the
expansion of Eq.~(\ref{Bexpand}) and collect terms up to second order in the
coordinates, 
\beq
U=u_i u_i+2u_i g_{ij}x_j+ (u_i c_{ijk}+ g_{ij}g_{ik}) x_j x_k +\mbox{h.o.t.} \label{Uexpand}
\eeq
For a stationary point in $x_i=0$ we set $\partial_p U=2u_i g_{ip}=0$. A trivial
solution is that the field is zero, $u_i=0$. Note however that this is generally only a stationary
point of $U$, not of $B$. For stationary points at
nonzero field, such as the minimum in a Ioffe-Pritchard trap, we must require
that $g_{ip}$ has an eigenvalue zero, i.e. 
\beq
\mbox{det}(g_{ij})=0. \label{detg0}
\eeq
Furthermore for a stationary point to occur, the field $u_i$ must be
parallel to the eigenvector of $g_{ip}$ with eigenvalue zero. In the
case of a IP trap this direction is usually called the axial direction.
The above condition, Eq.~(\ref{detg0}), expresses the fact that a IP trap
requires a point where the magnetic field locally looks like a
cylindrical quadrupole field, and that the axis of the quadrupole is the
trap axis. 

Since the gradient is a function of the spatial coordinates, $g_{ij}=g_{ij}(\mathbf{r})$,
the condition of Eq.~(\ref{detg0}) defines a
two-dimensional curved surface which we shall call the manifold $\cal{M}$. The points on the manifold are those
points in space where a stationary point for $B$ can be created by
choosing $u_i$ along the zero eigenvector of $g_{ij}$. Some examples of such manifolds in 
situations of practical interest are shown in Sec.~\ref{sec:practical}.

\subsection{Absence of field maxima in empty space}

If the condition $u_i g_{ip}=0$ is fulfilled, Eq.~(\ref{Uexpand}) is
simplified (up to second order) to:
\beq
U= u_i u_i+ t_{jk} x_j x_k, \label{Usimplified}
\eeq
where we defined
\beq
t_{jk}=u_i c_{ijk}+ g_{ij}g_{ik}.
\eeq
Note that the tensor $t_{jk}$ is symmetric. Its trace is a sum of squares, and therefore always
non-negative:
\beq
t_{kk}=g_{ik}g_{ik}\geq 0, \label{positivetrace}
\eeq
where we used the vanishing of partial traces,
Eq.~(\ref{zeropartialtrace}). We note that this gives the well known result that $U$
cannot have a local maximum in empty space, since a maximum would imply three
negative eigenvalues of $t$ and thus a negative trace. The absence of
maxima in empty space is known as Wing's theorem \cite{Win84} and is
here retrieved by a different route. Note that the non-negative trace
relies on $\mathbf{B}$ being irrotational and divergence free. Therefore the same conclusion
holds for electrostatic fields in vacuum. On the other hand
time-dependent $\mathbf{E}$ and $\mathbf{B}$ fields are not irrotational
and in fact do allow for a maximum of field magnitude in empty space
\cite{CorMonWie91, VelBetMei05}. 

\subsection{Trapping frequencies}

The potential for an atom in a magnetic field is given by:

\begin{equation}
\label{eqPot}
V=m_Fg_F\mu_B\sqrt{U},
\end{equation}
with $m_F$ the magnetic quantum number, $g_F$ the Land\'e factor and $\mu_B$ 
the Bohr magneton. If we make a harmonic approximation around the potential 
minimum we find that the trap frequencies are given by:

\begin{equation}
\label{eqTrapfreqs}
\omega_n=\sqrt{\frac{m_F g_F \mu_B T_n}{m \, u}},
\end{equation}
with $u=\sqrt{u_i u_i}$, with $T_n$ the eigenvalues of $t_{jk}$ and $m$ the mass of the atom.
Here we assume that the eigenvalues $T_n\geq 0$.

Combining this expression with Eq.~(\ref{positivetrace}) we find 
\begin{equation}
\omega_1^2+\omega_2^2+\omega_3^2\propto \frac{1}{u}\,g_{ik}g_{ik}.
\end{equation}
Thus, remarkably, we find that this combination of trap frequencies is independent of the
curvature $c_{ijk}$ and depends only on the gradient $g_{ij}$ and the uniform 
field $u_i$.

\subsection{Where can a IP trap be created?}

Having established that stationary points, including IP traps, can only be found on 
the manifold $\cal{M}$, we now address the question whether a IP trap can be created in any
arbitrary point on the manifold $\cal{M}$. For convenience we choose a coordinate frame that
diagonalizes $g_{ij}$, such that the zero eigenvector lies along
coordinate direction $\hat{e}_3$. The gradient tensor then takes a
very simple form, with $g_{11}=-g_{22}=a$ as the only nonzero components.
Since $u_i$ must be chosen along the zero eigenvector, we write
$\mathbf{u}=u_3\,\hat{e}_3$. We can then write the tensor $t_{jk}$ as 
\beq
(t_{jk})=\left( \begin{array}{ccc}
a^2 & 0 & 0\\
0 & a^2 & 0\\
0 & 0 & 0 \end{array}\right) +  u_3 \, c_{3jk}.
\eeq
Thus $t_{jk}$ depends on a single parameter $u_3$ that is a multiplier of
the symmetric and traceless tensor $c_{3jk}$. 

We can easily see the qualitative behavior of the eigenvalues of
$t_{jk}$ as a function of $u_3$. Obviously, for $u_3=0$ the eigenvalues
are $(T_1,T_2,T_3)=(a^2,a^2,0)$. For small values of $u_3$ we can use
perturbation theory to obtain the lowest eigenvalue to first order,
yielding 
\beq
T_3 \approx u_3 \, c_{333}.
\eeq
This shows that $T_3$ can be made either positive or negative by
choosing the sign of $u_3$. For small values of $u_3$ the other two
eigenvalues will remain positive. This means that for small $u_3$ the
stationary point will be either a IP trap or a saddle point with Morse
index of 1, Morse index being the number of negative eigenvalues. 
Note that we can only make a IP trap if $c_{333}\neq 0$. In fact, for the 
case that $c_{333}=0$ a counterexample is easily found.

For large enough positive or negative values of $u_3$ the term $c_{3jk}$
will become the dominant term. Since $c_{3jk}$ is traceless, it has
signature $(+,+,-)$ or $(+,-,-)$. Therefore for large values of $u_3$ we
will always have a saddle point. The sign of $u_3$ will determine whether
the Morse index is 1 or 2. 

Since for small $u_3$ the eigenvalues of $t_{jk}$ are given by $(a^2+u_3 c_{113},a^2+u_3 c_{223},u_3 \, c_{333})$,
we can tune the two non axial trap frequencies with $u_3$ but the
axial frequency is fixed by $c_{333}$ [Eq.~(\ref{eqTrapfreqs})].

\section{Moving stationary points}
\label{sec:MovingFieldMinima}

We now investigate how stationary points can be moved over the manifold 
by changing the uniform field $u_i$. 
We consider the situation that the spatial dependence of the magnetic
field is defined, e.g., by a configuration of permanent magnets. We can
influence the magnetic field pattern by applying a uniform external
field. In terms of the above quantities, $g_{ij}$ and $c_{ijk}$ are
fixed, $u_i$ is our control parameter.

\subsection{Moving Ioffe-Pritchard traps}

The movement of IP traps, which must clearly be constrained to the manifold, is important in applications that require trapped atoms to move, such as beam splitters and conveyor belts \cite{HanReiHan01}. We now calculate a displacement tensor $d_{jq}\equiv\partial x_j/\partial u_q$ in a stationary point $x_j=0$ on the manifold. This tensor describes how the position of a stationary point moves when the uniform field is changed. 
To calculate it, we use the condition for a stationary point, $\partial_p U = 0$, with $U$ as in Eq.~(\ref{Uexpand}), 
\begin{equation}
\partial _p U = 2 u_i g_{ip} + 2 (u_i c_{ijp}+g_{ij}g_{ip}) x_j=0.
\end{equation}
Taking the derivative with respect to $u_q$ and setting $x_j=0$, we solve for $d_{jq}$ and find:

\begin{equation}
d_{jq}\equiv\frac{\partial x_j}{\partial u_q}=-(t^{-1})_{jp}g_{pq} \label{moveIP}
\end{equation}

where $(t^{-1})_{jp}$ denotes the inverse tensor of $t_{jp}$. In the basis
where $g_{pq}$ is diagonal we see that $d_{j3}=0$, i.e., a small field in the axial
direction of a IP trap will not displace it. 
Since this means that the eigenvalue $D_3$ is zero, $d_{jq}$ is singular. Therefore $d_{jq}$ is a
mapping of a three-dimensional vector $u_q$ onto a two-dimensional space, namely the tangential plane to the manifold. It is spanned by the two eigenvectors $(\mathbf D_1$,$\mathbf D_2)$ 
corresponding to the nonzero eigenvalues. 
The vector $\mathbf D_1\times \mathbf D_2$ 
is normal to the manifold and is found to be proportional to
$(c_{133}, c_{233}, c_{333})$.

Note that 
during the movement the radial trap frequencies can be controlled using a bias field in the
axial direction of the IP trap [Eq.~(\ref{eqTrapfreqs})].

\section{The manifold and field zeroes}
\label{sec:TheManifoldAndFieldZeroes}

The manifold $\cal{M}$ is not only a powerful concept in the description of 
stationary points, it also has significance in the occurrence and movement of 
field zeroes. Such points of zero field are also minima of $B$ and can thus serve 
as atom traps. However, these so-called quadrupole traps (QT) suffer from higher 
trap loss rates due to Majorana spin flips near the region of zero field. 
The movement of QTs is important in loading procedures, i.e. to transport atoms into the final IP trap. 
In this section we investigate how field zeroes can be moved and what happens when they approach the manifold.

The manifold is the boundary between two regions of space $V^+$, $V^-$,
where $\det (g_{ij})>0$ and $<0$ respectively. Since $g_{ij}$ is traceless, and 
$\det (g_{ij})$ is the product of the eigenvalues, $g_{ij}$
must have two negative and one positive eigenvalues in $V^+$. Similarly,
in $V^-$ it has one negative and two positive eigenvalues. It is
impossible to move between the two regions $V^+$, $V^-$ without one of
the eigenvalues going through zero. 
This means that the manifold imposes restrictions on the movement of 
field minima that have a field zero. 

\subsection{Moving quadrupole traps}

Since a quadrupole trap is a zero of the magnetic field, it is
straightforward to give a prescription for how to move it by applying an
external field. We call the stationary field
$\mathbf{B}_\mathrm{stat}(\mathbf{r})$ and the desired trajectory
$\mathbf{r}(t)$. In order to move the field zero along the trajectory,
all we need to do is to use the external field
$\mathbf{B}_\mathrm{ext}(t)$ to cancel the local magnetic field, 
\beq
\mathbf{B}_\mathrm{ext}(t)=-\mathbf{B}_\mathrm{stat}(\mathbf{r}(t)).
\eeq

Like for the movement of stationary points, we can also express the movement of 
zeroes in terms of a displacement tensor 
$\partial x_j/\partial u_q$, Eq.~(\ref{moveIP}). 
For field zeroes this tensor takes a very simple form,
\begin{equation}
d_{jq}=\frac{\partial x_j}{\partial u_q}=-(g^{-1})_{jq}.
\end{equation}
Thus, generally speaking, field zeroes do not disappear when the uniform field $u_j$ is changed, 
they simply move through 3D space.

The situation is different when a field zero approaches the manifold between the 
regions $V^+$, $V^-$. On the manifold $\det (g_{ij}) = 0$ so that 
$d_{jq}$ does not exist. In fact, since one of the eigenvalues of $g_{ij}$ vanishes, the 
magnetic confinement along one direction of the magnetic QT vanishes.

\subsection{Pairs of zero field}

Having established how field zeroes (QTs) and IP traps move, the natural question arises what happens when a field zero approaches the manifold. We have just noted that field zeroes can be broadly categorized in two types, according to whether the signature of the gradient is $(+,+,-)$ or $(+,-,-)$. It turns out that the approach of the manifold by a field zero is accompanied by the approach by another field zero, of the opposite type, from the other side of the manifold. 

If a zero is sufficiently close to $\cal{M}$, we first assume that we can choose a point $x_i=0$ on $\cal{M}$ such that the position of the zero $x_i=\xi_i$ is in the direction $\mathbf{G_3}$ of the local quadrupole axis. 
Thus we have $\xi_1=\xi_2=0$ and $\xi_3\neq 0$. Furthermore $g_{i3}=0$ on $\cal{M}$ so that the requirement of zero field in $\xi_i$ simplifies to
\begin{equation}
u_i+c_{i33}\xi_3^2=0. \label{zeropairs}
\end{equation}
This shows immediately that the replacement $\xi_3\rightarrow -\xi_3$ yields another zero. The two zeroes are symmetrically placed around the point on $\cal{M}$, in the direction of the $\mathbf{G_3}$ axis. The zeroes must be of opposite type, since they are on opposite sides of $\cal{M}$. Finally, we note that the solution for $\xi_3$ of the above Eq.~(\ref{zeropairs}), namely $\xi_3=\pm\sqrt{-u_i/c_{i33}}$ implies that the local field direction $u_i$ must be proportional to $c_{i33}$. This is just the normal vector to the manifold as mentioned above, so the local field $u_i$
is normal to the manifold $\cal{M}$.

The choice of a point $x_i=0$ on $\cal{M}$ such that $\xi_i\parallel\mathbf{G_3}$ is possible unless 
$\mathbf{G_3}$ is parallel to the manifold. In this special case, where $c_{333}=0$, a zero arriving on the manifold can transform into a line of zero field, which lies entirely on the manifold. Such lines of zero require a treatment 
that goes beyond the second order field expansion. Evidence from numerical examples and some highly symmetric analytical examples suggests that such lines are closed loops on the manifold. Small perturbing fields can split this loop into a number of zero pairs, which can be macroscopically separated.

Finally we can also address the question whether the merging of field zeroes must necessarily take place on 
$\cal{M}$. We assume that two field zeroes are sufficiently close so that it is sufficient to expand the field upto second order. We define local coordinates $x_i$ around the first zero and $x'_i$ around the other. For the field expansion we can then write
\begin{equation}
g_{ij}x_j+\frac{1}{2}c_{ijk}x_jx_k=g'_{ij}x'_j+\frac{1}{2}c'_{ijk}x'_jx'_k
\end{equation}
Note that both $u_i$ and $u'_i$ are zero because we have two field zeros. We introduce the separation vector $\xi_i$ and substitute $x_i=\xi_i+x'_i$. Equating terms of equal powers in $x'_i$ we then obtain $(g_{ij}+g'_{ij})\xi_j=0$ and thus $\det(g_{ij}+g'_{ij})=0$. 
Thus, in the second order field expansion, the midway point between the two zeroes $x_j$ and $x'_j$ 
lies on the manifold. Furthermore, the two field zeroes must be of opposite type. 
Apparently, in order for field zeroes of {\em similar} type to approach each other 
the leading term in the field expansion must be higher than second order.

\begin{figure}[t!]
\includegraphics[width=90mm]{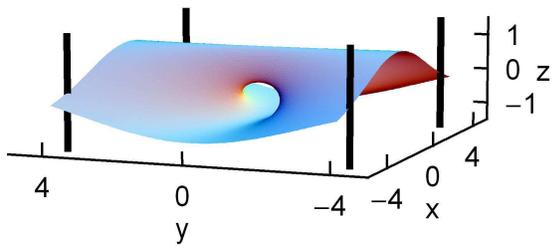}\\
\caption{(Color online) Manifold corresponding to the standard Ioffe-Pritchard trap, described by the field of Eq.~(\ref{eqStandardIP}) using $\epsilon=1$.
The black bars indicate the orientation of the Ioffe bars.}
\label{figStandardCurt}
\end{figure}

\section{Practical Ioffe traps}
\label{sec:practical}

In this section we consider the shape of the manifold $\mathrm{det}(g_{ij})=0$ in some cases of practical interest.

\subsection{Standard Ioffe-Pritchard trap}

The prototypical IP trap \cite{GotIofTel62,pri83} consists of four long current carrying wires (``Ioffe bars''), for
creating a cylindrical quadrupole field, in combination with two pinch coils 
creating confinement in the axial direction.
The field at the IP can be approximated by:
\begin{equation}
\label{eqStandardIP}
\mathbf{B}= \left(\begin{array}{c}
0\\
0\\
u\\
\end{array}\right)+a\left(\begin{array}{c}
x\\
-y\\
0\\
\end{array}\right)+
\frac{c}{2}\left(\begin{array}{c}
-xz\\
-yz\\
z^2-\frac{1}{2}(x^2+y^2)\\
\end{array}\right)
\end{equation}
Where $u$, $a$ and $c$ are the uniform field, radial gradient and axial curvature, respectively.
The manifold produced by this field is shown in Fig.~\ref{figStandardCurt}.
It is described by the equation:

\begin{equation}
\epsilon (x^2-y^2)+z(2z^2+y^2+x^2-2\epsilon^2)=0
\end{equation}

Where $\epsilon=2a/c$. Note that for $\epsilon=0$ this describes the flat surface
$z=0$. The hole in the manifold has typical size $\epsilon$. In the point $(0,0,0)$ the eigenvectors of $d_{jq}$ point in the $x$, $y$, and $z$ directions and so the lowest order displacement of the IP under influence of an external field is in a direction prependicular to the axial direction. The shape shown is only realistic in the region where Eq.~(\ref{eqStandardIP}) is a good approximation of the field.

\subsection{Z-wire Ioffe trap}

\begin{figure}[t!]
\includegraphics[width=85mm]{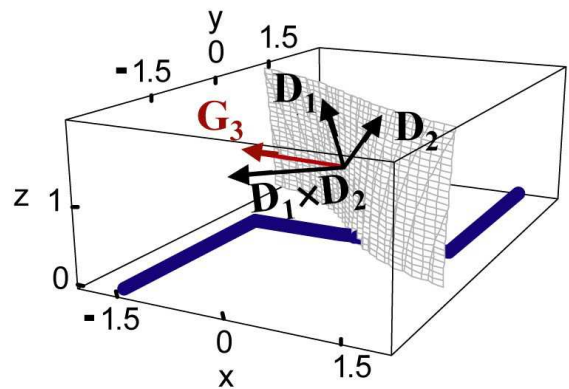}\\
\caption{(Color online) Manifold created by a Z-shaped wire. Since the manifold is given by the gradient its shape
does not depend on the bias field. The current is only a multiplier and
is also of no importance for the shape. The vectors $\mathbf{D_1}$,
$\mathbf{D_2}$, $\mathbf{G_3}$ and $\mathbf{D_1}\times \mathbf{D_2}$ have been drawn in
the point (0,0,1). The vector $\mathbf{G_3}$ gives the axial direction of
the IP, $\mathbf{D_1}\times \mathbf{D_2}$ is normal to the manifold.}
\label{figZshapeCurt}
\end{figure}

\begin{figure}[t!]
\includegraphics[width=80mm]{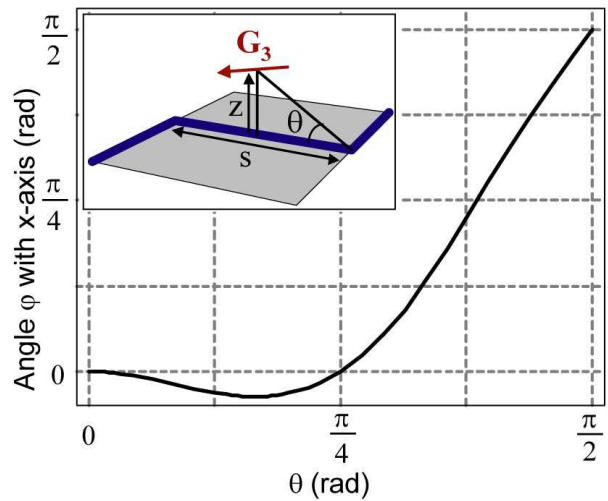}\\
\caption{(Color online) The azimuthal angle $\varphi$ of $\mathbf{G_3}$ with respect to the $x$-axis as a function of $\theta$. Note that directly above the middle of the central wire $\mathbf{G_3}$ always lies in the $xy$-plane. The inset shows the geometrical meaning of $\theta$.}
\label{figAngle}
\end{figure}

A method routinely used in atom chip experiments for creating IP traps
involves a current carrying wire bent in a Z-shape in combination with
a uniform bias field \cite{ReiHanHan99}. In Fig.~\ref{figZshapeCurt} the manifold for such a 
wire is shown together with the eigenvectors of $d_{jq}$ and the IP axis $\mathbf{G_3}$. The vector
$\mathbf{G_3}$ straight above the middle of the central wire always lies in the $xy$-plane. We find that 
the azimuthal angle $\varphi$ between $\mathbf{G_3}$ and the $x$-axis is given by: 
\beq
\varphi=-\arctan\left(\frac{\cos2\theta \sec\theta}{2+\cot^2 \theta}\right)
\eeq

Here $\tan \theta = 2z/s $, where $s$ is the separation between the two wires in the $y$-direction and $z$ is the height above the central wire. This result is plotted in Fig.~\ref{figAngle}. For $z=0$, $\mathbf{G_3}$ points in the  $x$-direction, in the limit $z\rightarrow\infty$ it points in the $y$-direction. For a given $s$, 
$\mathbf{G_3}$ points in the $x$-direction if we choose $z=\frac{1}{2}s$.

\subsection{Array of Ioffe traps}

\begin{figure}
\includegraphics[width=80mm]{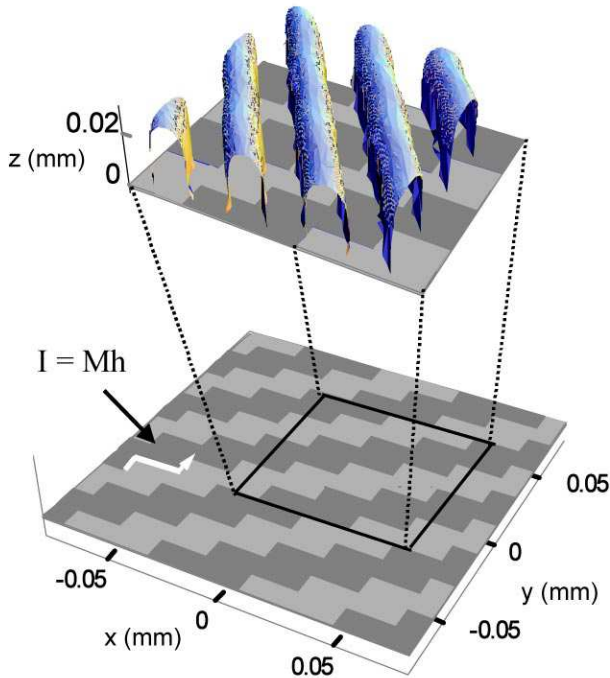}
\caption{(Color online) Array of magnetic material (dark regions). The magnetization is out of
plane. The equivalent edge current forms ``z''-shapes at every lattice point. In combination
with a bias field in the $y$-direction IP traps can be created. The shape of the manifold is shown in the inset.}
\label{arraysinone}
\end{figure}

The exact shape of the manifold is of particular importance in arrays of IP traps
that might be used as conveyor belts/shift registers. Such devices are promising for quantum information
processing applications \cite{GhaKieHan06}. Atoms sitting at a lattice site that is connected to another via the manifold
can be shifted there using a uniform bias field, while remaining a IP trap. Moreover,
the trap frequencies along the way can be tuned using Eq.~(\ref{eqTrapfreqs}). 

To make these ideas more explicit, we discuss an array of IP traps created
by permanent magnetic material in combination with a uniform bias field. 
In Fig.~\ref{arraysinone} the array of magnetic material is shown. Since its magnetization
is out of plane we can think of the material as having an equivalent current of
magnitude $M h$ running around its edges, where $M$ is the magnetization and $h$ is
the height of the material. This equivalent current forms a Z-shape at every
lattice site in the array. We apply a uniform bias field in the $y$-direction 
to create Ioffe traps at all lattice sites. 

\begin{figure}
\includegraphics[width=70mm]{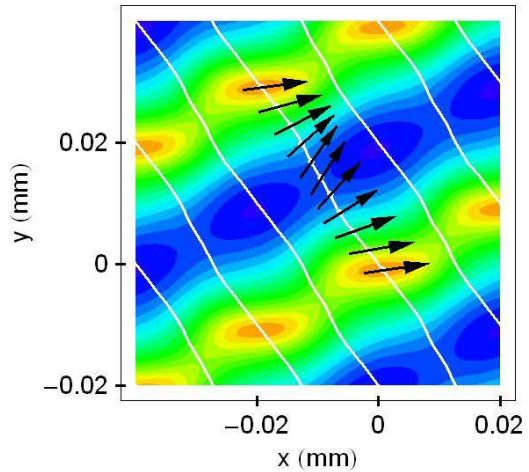}
\caption {(Color online) Two-dimensional cross-section showing the magnetic potential
and the manifold at $z=10\mu m$. The IP traps can be transferred over the manifold. For one IP the route
to the neighbouring lattice site is shown. The arrows indicate how the trap axis varies
along the way.}
\label{figPotcur}
\end{figure}

In this particular design the magnetization is $M=$800~kA/m and the height of the
material is $h=$250~nm. A bias field
of 17.9~G in the $y$-direction then produces IP traps at 10~$\mu$m  
from the surface with trap frequencies
(21, 20, 5.4) kHz and a residual field of 4.5~G at the trap bottom.

We are interested in where the individual IP traps can be moved. Therefore
we draw the manifold over a region of the array containing several unit cells.
As can be seen in Fig.~\ref{arraysinone}, IP traps are only connected in a diagonal direction from lower right to upper left.
This means that this array can be used as a shift register only in this direction.
The manifold clearly does not allow shifting the IP traps in the perpendicular direction. 
One could try to move the traps in the perpendicular 
direction as field zeroes (QT). However, we find that this leads to a sequence of splitting 
and recombination of zeroes, every time the manifold is crossed. 
Thus, in spite of its appearance, this is a 1D shift register. 

Finally, for the sake of completeness, we show in Fig.~\ref{figPotcur} how 
the axial direction of the Ioffe traps varies over the manifold.

\section{Conclusion}

In conclusion, we have shown that Ioffe Pritchard traps as well as other stationary points of $B(\mathbf{r})$
are confined to a two-dimensional curved manifold $\cal{M}$ defined by $\det (g_{ij})\equiv\mathrm{det}(\partial B_i/\partial x_j)=0$. 
Furthermore, in any point of $\cal{M}$ where the local quadrupole axis is not parallel to $\cal{M}$ a IP trap or other stationary point can be created by choosing
the magnetic field parallel to this axis, i.e. the eigenvector of $g_{ij}$ corresponding to eigenvalue zero. 
We have given an expression for the movement of stationary points over the manifold, in response to a change 
of an external uniform control field. 

We have shown that the same manifold plays an important role in the behavior of field zeroes, and separates 
regions of space where field zeroes are of an opposite type. These field zeros can be moved anywhere in their 
respective region of space but can only disappear on the manifold. In this annihilation two field zeros of 
opposite type merge and form a Ioffe Pritchard trap. Similarly, a IP on the manifold can be made to split into 
two field zeroes of opposite type, on opposite sides of the manifold.

The manifold is a new conceptual tool in designing magnetostatic or electrostatic potentials. Understanding how IP traps can be moved over this manifold is of great use in experiments that require some level of dynamics such as shift registers for trapped atoms or double well experiments. The shape of the manifold is important in loading protocols for permanent magnetic atom chips, defining possible routes for transfering Ioffe Pritchard traps. 
%\vspace{2em}

\begin{acknowledgments}

We gratefully acknowledge helpful discussions with
Mikhail Baranov and Thomas Fernholz. This work is part of the research program
of the Stichting voor Fundamenteel Onderzoek van de Materie (Foundation
for the Fundamental Research on Matter) and was made possible by
financial support from the Nederlandse Organisatie voor Wetenschappelijk
Onderzoek (Netherlands Organization for the Advancement of Research). It was also supported by the EU under contract MRTN-CT-2003-505032.

\end{acknowledgments}

\end{document}